\documentclass[journal=nanoletter,manuscript=article]{achemso}

\usepackage{chemformula} 
\usepackage[T1]{fontenc} 

\usepackage{graphicx} 
\usepackage{subcaption} 
\usepackage[separate-uncertainty = true, multi-part-units = single, list-units = single, range-units = single]{siunitx} 
\DeclareSIUnit\Molar{\textsc{m}}
\DeclareSIUnit\rpm{rpm}
\DeclareSIUnit\ppm{ppm}

\DeclareSIUnit\kbt{k_BT}

\usepackage[noabbrev,nameinlink]{cleveref}
\usepackage{multirow}
\usepackage{makecell}
\usepackage{colortbl}

\title{Wrapping pathways of anisotropic dumbbell particles by giant unilamellar vesicles}
\author{Ali Azadbakht}
\affiliation[Leiden University]{Soft Matter Physics, Huygens-Kamerlingh Onnes Laboratory, Leiden University, PO Box 9504, 2300 RA Leiden, the Netherlands}
\altaffiliation{These authors contributed equally to this work.}

\author{Billie Meadowcroft}
\affiliation[Austria]{Institute of Science and Technology Austria, 3400 Klosterneuburg, Austria}
\alsoaffiliation[UCL1]{Department of Physics and Astronomy, Institute for the Physics of Living Systems, University College London, London WC1E 6BT, United Kingdom}
\altaffiliation{These authors contributed equally to this work.}

\author{Thijs Varkevisser}
\affiliation[Leiden University]{Soft Matter Physics, Huygens-Kamerlingh Onnes Laboratory, Leiden University, PO Box 9504, 2300 RA Leiden, the Netherlands}
\alsoaffiliation[UVA]{Van der Waals-Zeeman Institute, Institute of Physics, University of Amsterdam, Science Park 904, 1098 XH Amsterdam, Netherlands}
\altaffiliation{These authors contributed equally to this work.}

\author{An\dj ela \v{S}ari\'c}
\affiliation[Austria]{Institute of Science and Technology Austria, 3400 Klosterneuburg, Austria}
\author{Daniela J. Kraft}
\affiliation[Leiden University]{Soft Matter Physics, Huygens-Kamerlingh Onnes Laboratory, Leiden University, PO Box 9504, 2300 RA Leiden, the Netherlands}
\email{kraft@physics.leidenuniv.nl}

\begin{tocentry}
\centering
\includegraphics []{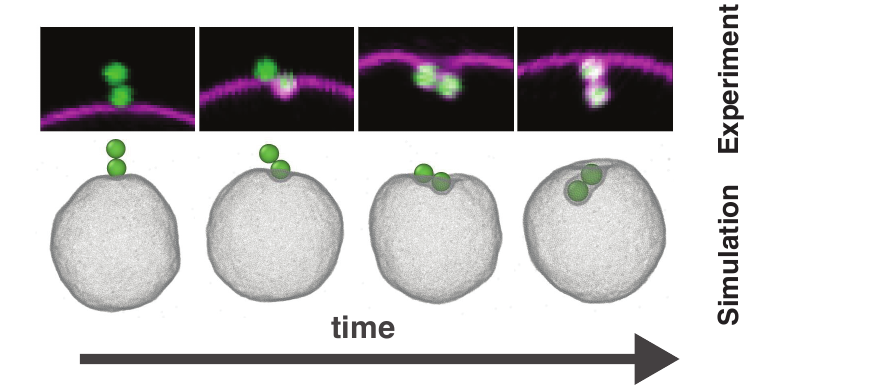}
\end{tocentry}

\begin{document}
\maketitle
\begin{abstract}
Endocytosis is a key cellular process involved in the uptake of nutrients, pathogens or the diagnosis and therapy of diseases.
Most studies have focused on spherical objects, whereas biologically relevant shapes can be highly anisotropic. 
In this letter, we use an experimental model system based on Giant Unilamellar Vesicles (GUVs) and dumbbell-shaped colloidal particles to mimic and investigate the first stage of the passive endocytic process: engulfment of an anisotropic object by the membrane. 
Our model has specific ligand-receptor interactions realized by mobile receptors on the vesicles and immobile ligands on the particles.
Through a series of experiments, theory and molecular dynamics simulations, we quantify the wrapping process of anisotropic dumbbells by GUVs and identify distinct stages of the wrapping pathway. We find that the strong curvature variation in the neck of the dumbbell as well as membrane tension are crucial in determining both the speed of wrapping and the final states. 
\end{abstract}

The engulfment of objects through the cell membrane is critical for endocytic processes such as phagocytosis \cite{Douglas2006,Manchester2006,Lewis2006} and receptor-mediated endocytosis. The latter is often exploited by viruses for cell entry and proliferation \cite{schmidt2012poxvirus} and key to nanomedical applications such as drug delivery and imaging\cite{mitchell_engineering_2021}. 
To single out receptor-mediated effects from active mechanisms involved in the engulfment\cite{Richards2014}, simplified passive model systems can be employed, which recently led to a conclusive understanding of the wrapping of spherical objects\cite{Spanke2020,Spanke2022}. However, biological objects such as bacteria and viruses \cite{young_selective_2006,levy_virology_1994,schmidt2012poxvirus} as well as nanoparticles relevant for applications in nanomedicine but also nanotoxicology \cite{cho_therapeutic_2008} often posses non-spherical shapes. Moreover, \textit{in vitro} experiments with nanoparticles and simulations have shown that the size and shape influence their likelihood to be taken up by endocytosis 
\cite{Richards2014,Aoyama2003,Richards2016,Chithrani2006,Chithrani2007,Li2013,Dasgupta2014}.

The wrapping pathways of spheres at sufficiently low membrane tensions have been shown to be a continuous transition from attached to fully wrapped, occurring either spontaneously or after activation  \cite{VanDerWel2016,Spanke2020,Spanke2022}. In contrast, anisotropic particles such as ellipsoids and rods, are expected to reorient during the wrapping process or become trapped in metastable states due to their varying curvature.\cite{decuzzi_receptor-mediated_2008, Vacha2011, Bahrami2013OrientationalMembranes, yang_influence_2013, Huang2013, dasgupta_shape_2014, Bahrami2014, Tang2018, Agudo-Canalejo2020} 
The aspect ratio of these particles as well as the degree of rounding of their tip were the key parameters affecting the wrapping orientation with respect to the membrane and their metastable and stable states\cite{dasgupta_shape_2014, Agudo-Canalejo2020}. Despite the extensive work in theory and simulations and exciting observations on shape-dependence in phagocytosis\cite{Champion2006}, no experimental work has investigated the passive wrapping process of anisotropic particles by lipid membranes and tested these predictions yet.

In this letter, we employ an experimental model system based on Giant Unilamellar Vesicles (GUVs) and colloidal dumbbell particles to investigate the wrapping of micrometre-sized anisotropic objects by lipid membranes. Our model system is designed to have mobile ligands on the vesicles and immobile receptors on the particles mimicking receptor-mediated endocytotic systems \cite{VanDerWel2016, VanDerWel2017,Sarfati2016}. We quantify the wrapping pathways of anisotropic dumbbells by lipid membranes and test if their initial orientation affects the final states. Molecular dynamics simulations of the same system corroborate our experimental data, allowing us to inspect the dynamics of the process that was inaccessible to experiment. We find that the strong curvature variation in the neck of the dumbbell as well as membrane tension and not their initial orientation are crucial in both determining the speed of wrapping and the final states.

\begin{figure}[hbt!]
\centering
\includegraphics[width=1\linewidth]{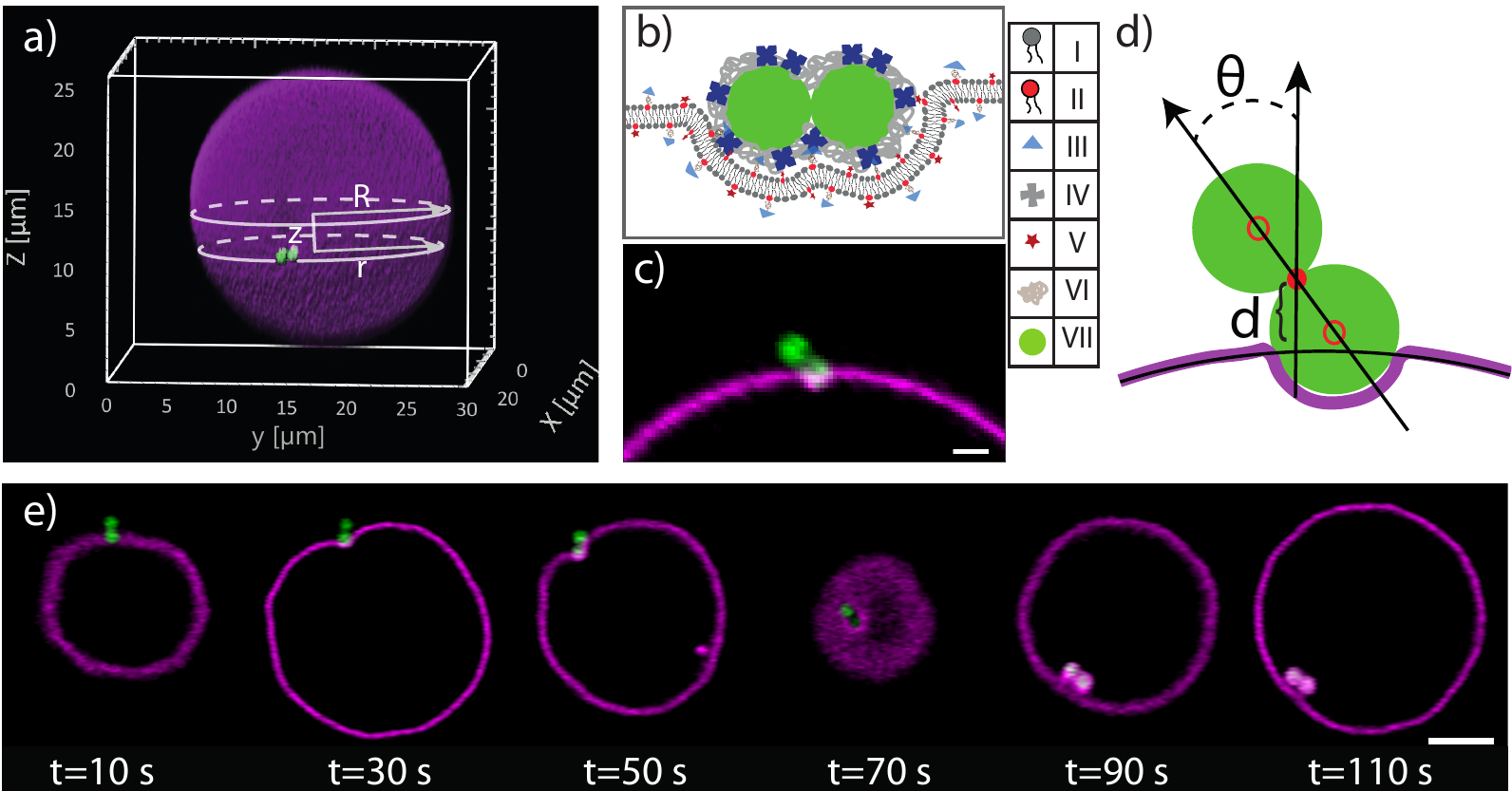}
\caption{\textbf{Experimental setup to quantitatively measure the wrapping process of a dumbbell colloid by a GUV} 
a) 3D confocal reconstruction of a GUV in magenta and a dumbbell particle in green with an indication of the relative height $z$ from the equator of the GUV, radius of GUV $R$, and cross section radius of the vesicle at the location of the dumbbell, $r$.
b) Detailed schematic of ligand-receptor based binding scheme between the dumbbell and GUV. I-DOPC lipid II-DOPE lipid III -Biotin IV -NeutrAvidin V-Rhodamine, VI -Polyethylene glycol (PEG) VII -Polystyrene particle. (Not to scale)
c) Representative confocal images reconstructed from two channels, (1) dumbbell excited by 488 nm laser light and emission collected between 500-550 nm  (2) GUV excited by 561 nm laser light and emission collected in 580-630 nm (scale bar 1$\mu m$).
d) Schematic representation of the parameters $d$ and $\theta$ used for the quantitative description of the wrapping process. 
e) Time series of snapshots of confocal images of a dumbbell being wrapped by a vesicle (scale bar 4$\mu m$).
}
\label{fig:Fig.1}
\end{figure}

We investigate the wrapping process of anisotropic objects by a lipid membrane using a model system consisting of GUVs and colloidal particles, (see Fig. \ref{fig:Fig.1}a). We chose the simplest object that features anisotropy: a dumbbell shaped colloidal particle that consists of two equal sized spheres. The colloid dumbbells were obtained from aggregating polystyrene spheres with diameter $d_s$=0.98$\pm$ 0.03 $\mu$m\cite{VanDerWel2017a} by briefly lowering the pH to 5.3 and then quenching the process by increasing the pH to 8.6 \cite{Meester2016ColloidalParticles}. This process yielded 5-10\% dimers with a long axis of 1.96 $\pm$ 0.06 $\mu$m and a short axis of 0.98 $\pm$ 0.03 $\mu$m. GUVs were prepared by electroswelling from 97.5\% w/w 1,2-dioleoyl-sn-glycero-3-phosphocholine (DOPC). 

To realize strong ligand-receptor mediated binding we doped the GUVs with 2\% w/w 1,2-dioleoyl-sn-glycero-3-phosphoethanolamine-N-[biotin-2000] (DOPE-PEG2000-Biotin) and the dumbbells with 2.2$\times$10$^3$/$\mu$m$^2$ NeutrAvidin following \cite{VanDerWel2017a}, see Fig. \ref{fig:Fig.1}b and c and see particle functionalization and quantification of binding affinity in Supporting Information. We suppress electrostatic interactions by working in 50 mM Phosphate Buffered Saline, and achieve colloidal stability by coating the dumbbells with polyethyleneglycol (PEG5000). 
Imaging of the position and orientation of the dumbbells and membranes in three dimensions was made possible by dying the colloids with BODIPY, represented by a green color throughout the manuscript, as well as including 0.5\% w/w 1,2-dioleoyl-sn-glycero-3-phosphoethanolamine-N-(lissamine rhodamine B sulfonyl) (DOPE-Rhodamine) into the GUVs, represented by a magenta color. See Fig. \ref{fig:Fig.1}c. 
Confocal stacks and image sequences were acquired with an inverted Nikon TI-e microscope, equipped with a 60x (NA 1.2) objective and A1-R scan head. 2D image sequences were taken at 59 fps, which enables tracking of the dumbbells in real time. Experimental details are described in the Supporting Information. 

To initiate the wrapping process, we used optical tweezers to bring dumbbell particles in contact with the GUV. They subsequently diffused on the GUV surface before suddenly and quickly becoming wrapped, a process that took between a few seconds and a few minutes depending on membrane tension, see Figure \ref{fig:Fig.1}e and Movie S1. To capture the wrapping process with high speed, we adjusted the focal height during acquisition of the image sequence. After wrapping, the dumbbell continued to diffuse on the inside of the vesicle. 

We quantify the wrapping process of a dumbbell by measuring the angle $\theta$ between the major axis of the dumbbell and surface normal of the GUV and distance $d$ of the dumbbell with respect to the undistorted surface of the GUV, see Figure \ref{fig:Fig.1}d. We inferred the 3D position of the dumbbell from the position of its lobes with respect to the GUV. To improve the accuracy of tracking, particles were tracked only when their center of mass was between -0.8$R$<$z$<0.8$R$, and when both lobes were in focus. Details are described in the Supporting Information.

\begin{figure}[hbt!]
\centering
\includegraphics[width=1\linewidth]{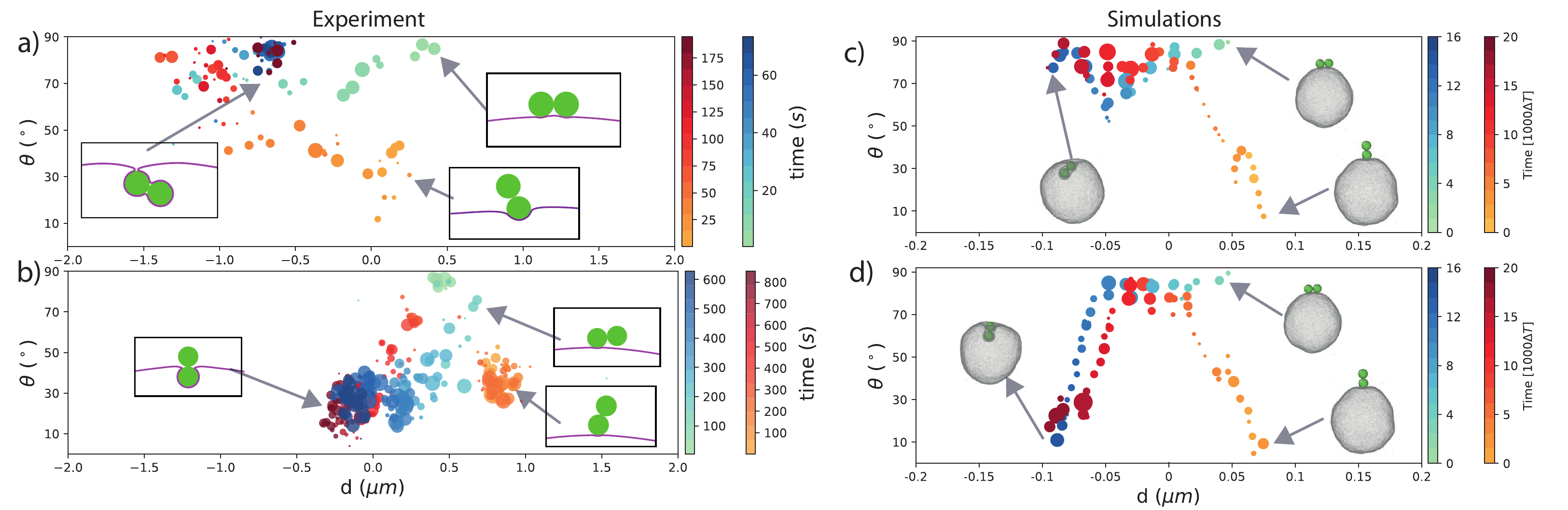}
\caption{\textbf{Quantitative wrapping pathway of dumbbell particles by for GUVs.} Tilt angle $\theta$ and distance $d$ of the dumbbell from the vesicle surface obtained from a,b) experiments and c,d) simulations as a function of time. In all panels, green-blue pathways indicate dumbbells starting from a vertical position with respect to the vesicle surface, and yellow-red pathways indicate dumbbells that initially start almost horizontally with respect to the membrane. Time is indicated by color, specified by colorbars for each panel. 
a) Experimentally obtained pathways for a dumbbell initially oriented parallel or perpendicular to the membrane surface to a  fully wrapped end state. Each data point represents an average over 1s.
b) Experimentally obtained pathways taken by a dumbbell initially oriented parallel or perpendicular to the membrane surface to the half-wrapped end state. Each data point represents an average over 5s.
c) Simulations of pathways for a dumbbell initially oriented parallel and perpendicular to the membrane surface to the fully-wrapped end state. This was the most common stable state with $\sim 90 \% $ of dumbbells reaching this end state. 
d) Simulation of pathways for a dumbbell initially oriented parallel and perpendicular to the membrane surface to the half-wrapped end state.
a-d) Circle size indicates the number of images used for the average. Simulation time is in expressed in $\Delta T = 0.01\tau_0$, $\tau_0$ being MD unit of time.
}
\label{fig:Quantitave_pathway}
\end{figure}

We show confocal microscopy snapshots of a typical wrapping pathway in Figure \ref{fig:Fig.1} e, and quantitative data of $\theta$ and $d$ for exemplary pathways in Figure \ref{fig:Quantitave_pathway}a and b. Surprisingly, we find that the dumbbells end up in one of two states even though they start from different initial orientations: (1) both lobes are either being fully wrapped (Fig. \ref{fig:Quantitave_pathway}a), or (2) a single lobe is being wrapped, such that the dumbbell is engulfed up to its waist by the membrane (Fig. \ref{fig:Quantitave_pathway}b). The green-blue points in Fig. \ref{fig:Quantitave_pathway}a and b represent dumbbells attached almost parallel to the membrane at the beginning of the process, whereas the yellow-red points represent dumbbells attached roughly perpendicular with respect to the membrane initially. Other starting orientations also lead to either a fully wrapped or a half wrapped dumbbell, but the probability for reaching either state was influenced by the initial position as we will discuss below.

If the dumbbell is oriented parallel to the membrane initially ($\theta\approx 90^\circ$ and proceeds to a fully wrapped state, then it tilts in the first part of the engulfment process to about 60$^\circ$. Subsequently, its CoM moves inward to almost $d\approx 1.5 d_s$ from the undisturbed membrane contour, before returning to a more parallel orientation and an insertion depth about $d\approx 0.7 d_s$. This overshooting and recoil is similar to that observed for spheres previously\cite{Dietrich1997,Spanke2022}.
If the dumbbell initially is roughly perpendicular the membrane, it first becomes oriented more precisely perpendicular until it is covered halfway ($d=0$ and $\theta\approx 10 ^\circ$) before being wrapped further and finally ending in a more parallel orientation at a similar distance from the undisturbed membrane as the initially parallel dumbbells. 

For final states where one lobe is being wrapped only, an initially perpendicular dumbbell first reorients more parallel before becoming engulfed until its waist while becoming perpendicular again. An initially parallel dumbbell proceeds to reorient perpendicular while being engulfed, see Fig. \ref{fig:Quantitave_pathway}b. The gap in the yellow-red trace at $\theta \approx 55^\circ$ and $d$=0.5 $\mu m$ was caused by the dumbbell going through an orientation that was filtered out for accuracy as described above. 

To obtain more quantitative results for the dynamics of the system we carried out coarse-grained (CG) molecular dynamics (MD) simulations of anisotropic dumbbell particles being wrapped by a membrane. Besides the advantage of easily measuring dynamic properties, in these simulations we are also able to control the size of the vesicle and dumbbell, the membrane tension and the interaction strength between dumbbell and membrane and thus probe a wider parameter space than is available to experiments. 

The membrane is modelled using a one particle thick fluid surface developed by Yuan et al \cite{Yuan2010} which reproduces the mechanical properties associated with biological membranes \cite{curk2018controlling}. Using this model, we simulate spherical membrane vesicles and change the membrane tension by the addition of small solute particles on the inside and outside of the vesicle \cite{vanhille2021modelling}. The solute particles only interact via volume exclusion and produce a pressure force when the inside and outside concentrations are different. The dumbbell colloid is then placed on the membrane in either a vertical or horizontal initial condition and due to the attractive interaction between the membrane beads and the dumbbell, the dumbbell is slowly wrapped and engulfed by the vesicle. Details can be found in the Supporting Information.

The results obtained from simulations show qualitatively similar behavior as in the experiments, see Figure \ref{fig:Quantitave_pathway}. Again, both final states, i.e. i) one lobe attached and ii) fully engulfed, could be reached from any initial position, and the pathway they took was influenced by the initial orientation. Interestingly, our simulations suggest that the initial position strongly influences the first part of the wrapping process and to a lesser degree the second half, which is observed to be similar for both extreme initial orientations. The observation that the wrapping pathways from different initial positions can result in the same final position shows that there is an energy minimum for the GUV-dumbbell system independent of the initial position of the dumbbell. 
In all observed pathways towards the fully wrapped state, the dumbbell particle tilts during the engulfment suggesting that this requires less bending energy. 

A similar reorientation upon wrapping was observed for linear aggregate of particles \cite{vsaric2012mechanism} and elongated ellipsoids \cite{Bahrami2013OrientationalMembranes, Bahrami2014, Tang2018, Agudo-Canalejo2020}. Ellipsoids have been found to become first adhered by the side, before rotating to the tip upon being wrapped by the membrane \cite{Bahrami2014}. For sphero-cylindrical particles that were initially touching with their tip, a rotation-mediated wrapping was also seen \cite{Huang2013,Dasgupta2014}, which can rotate the particle from a standing to a lying position at high aspect ratios. The first point of contact has been predicted to be crucial for the ultimate fate of a non-spherical particle\cite{Tang2018, Agudo-Canalejo2020}. In contrast, for the dumbbell particles used here rotation is not driven by a variation of particle curvature, but primarily by thermal fluctuations and possibly inhomogeneities in the ligand coating density, because of the constant curvature of the constituent spheres of the dumbbells. The only region of curvature variation is the dumbbell neck, which we will show to play a crucial role in the wrapping.

\begin{figure}[hbt!]
\centering
\includegraphics[width=1\linewidth]{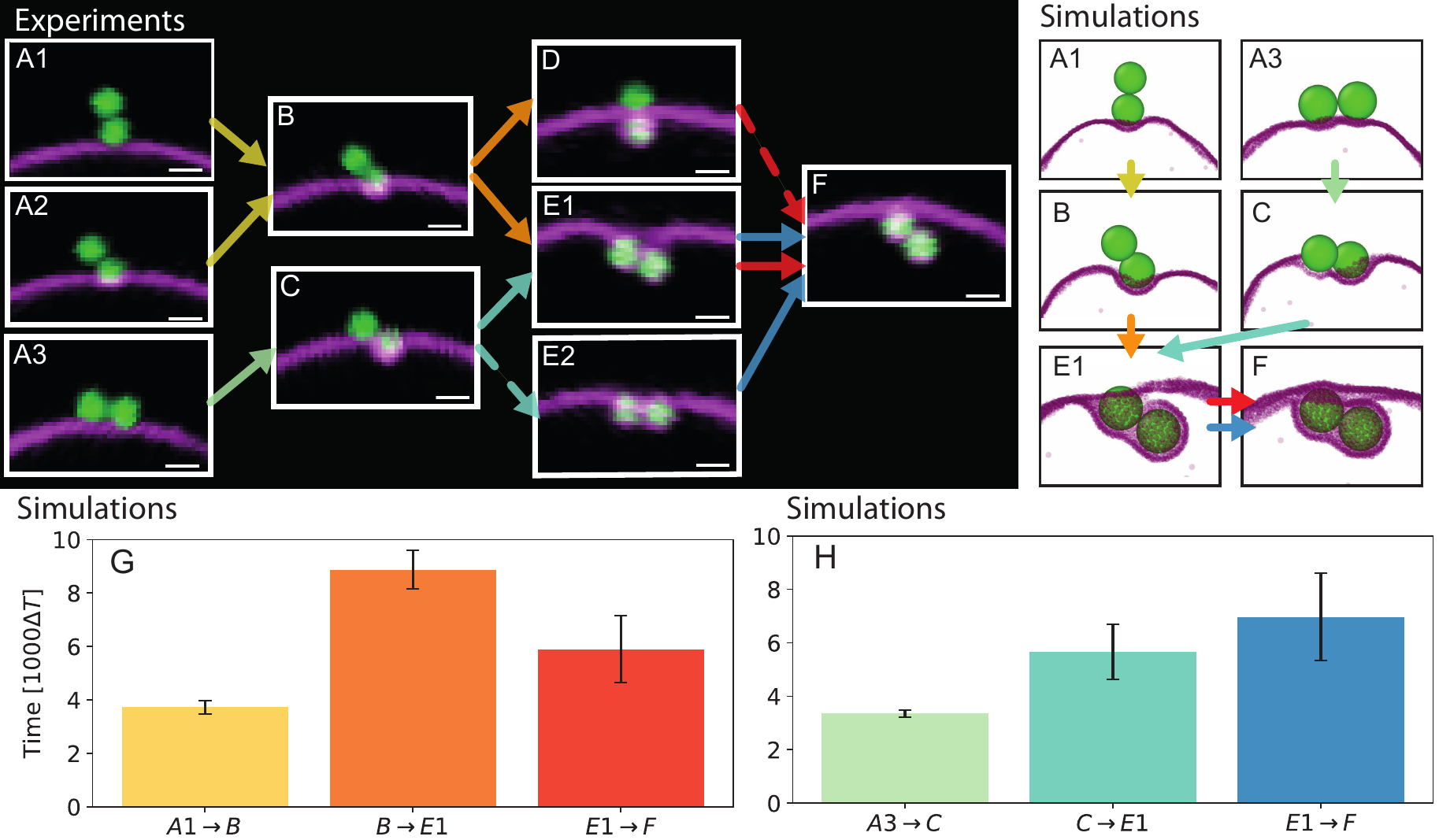}
\caption{\textbf{Overview of the observed wrapping pathways.} A1-F) Confocal images of the possible orientation of a dumbbell (All scale bars denote 1$\mu$m). Arrows indicate the directions of the possible wrapping pathways, and dashed arrows illustrate transitions that were rarely observed.
G) Measurements of the time between the states for the horizontal dumbbell starting position, given in simulation timesteps. 
H) Measurements of the time between the states for the vertical dumbbell starting position.}
\label{fig:Wrapping pathways illustrations}
\end{figure}

From the many wrapping processes we observed in experiments and simulations, we identified a number of key intermediate states during the engulfment that ultimately determined the final state. A decisive event during the wrapping of the first lobe is whether the second lobe gets bound to the membrane. This is always the case if the particle starts out being perfectly parallel and thus with both lobes attached (Figure \ref{fig:Wrapping pathways illustrations}A3). If the particle initially is attached with a single lobe (\ref{fig:Wrapping pathways illustrations}A1 and A2), however, tilting during the engulfment may attach the second lobe (\ref{fig:Wrapping pathways illustrations}B). In principle, since one lobe is spherical one may expect engulfment to proceed uniformly, not inducing or requiring any tilt. However, any inhomogeneity in the coating density of the ligands on the dumbbells, as well as thermal fluctuations will tilt the particle and may induce contact of the second lobe to the membrane. Since biotin-Neutravidin interactions are essentially irreversible at room temperature, attachment of the second lobe always precludes achieving a final state where only one lobe is wrapped. If the second lobe does not attach, the single-wrapped lobe state is reached (\ref{fig:Wrapping pathways illustrations}D). Otherwise, the dumbbell will wrap both lobes consecutively, either in a symmetric fashion (\ref{fig:Wrapping pathways illustrations}E2) or in an asymmetric way (\ref{fig:Wrapping pathways illustrations}E1), leading to the fully wrapped state. The symmetric wrapping is unstable, and eventually leads to Fig. \ref{fig:Wrapping pathways illustrations}F in which both lobes are covered. The angle the dumbbell makes with the membrane after wrapping completed can vary. In this end state, a small neck connected the fully wrapped dumbbell at one lobe with the vesicle, see Fig. \ref{fig:Wrapping pathways illustrations}F. 

To quantify the time evolution, we measured the transition times between the different wrapping states. Membrane tension was found to be crucial for the overall wrapping time, see below, and therefore simulations were used for quantitative measurements of the transition times and experiments for qualitative comparison. While the initial wrapping of the first lobe in the simulations is almost equally fast for the different initial states (see Fig. \ref{fig:Wrapping pathways illustrations}G and H), the wrapping slowed down significantly when the membrane was crossing the waist (Fig. \ref{fig:Wrapping pathways illustrations}G and H). This signifies an energy barrier stemming from the high bending energy required to adapt to the strong variation in curvature of the particle surface. For dumbbells with both lobes attached, we observed slowing down at the waist (Fig. \ref{fig:Wrapping pathways illustrations}G). For dumbbells attached with a single lobe only, the wrapping process stopped for a longer time at the waist (Fig. \ref{fig:Wrapping pathways illustrations}H). We observed the same qualitative behavior in experiments, both for tense and floppy GUVs, indicating that the bending energy required to continue wrapping largely exceeded the energy gained from adhesion. In experiments, in less than 10\% of the cases, we observed dumbbells wrapped with one lobe (\ref{fig:Wrapping pathways illustrations}D) to suddenly transition to the fully engulfed state within about 10 minutes, but never observed this within the timescales used in simulations in line with ref. \cite{Richards2016}. The high bending energy costs at the waist and the significantly faster wrapping for tilted dumbbells observed in both simulations and experiments suggest that wrapping a tilted dumbbell is less energetically costly than one that is oriented perpendicular to the membrane.\cite{Bahrami2014} The strong trapping at the waist also causes single-lobe wrapped dumbbells to attain their stable insertion depth $d$ without overshooting and recoil.

The probability of following a specific pathway and reaching one of the two final states as qualitatively observed in experiments, depended on two factors: the membrane tension of the GUV and the dumbbell's angle $\theta_0$ with respect to the membrane's surface normal during the initial wrapping. The higher the surface tension of the GUV, the more likely it was for the dumbbell to end up in situation \ref{fig:Wrapping pathways illustrations}D. Large fluctuations of the vesicle's surface enabled the dumbbell to attach to the non-wrapped lobe. The larger the angle $\theta$ in situation \ref{fig:Wrapping pathways illustrations}A2, and thus the closer to the membrane it started out at the more likely it was for the dumbbell to end up in situation \ref{fig:Wrapping pathways illustrations}B and hence E1.

\begin{figure}[hbt!]
\centering
\includegraphics[width=1.\linewidth]{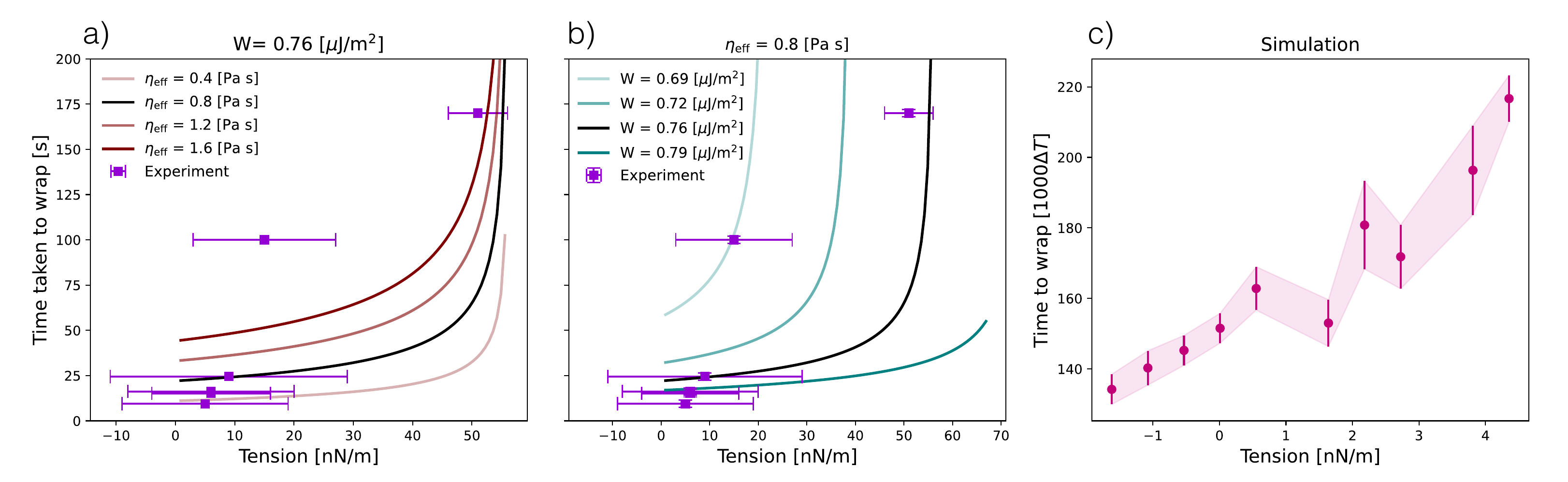}
\caption{\textbf{Measurement of the time required to fully wrap a dumbbell-shaped particle as a function of membrane tension} (a) Experimental data (points) and theoretical predictions (lines) for different membrane viscosity in the range of 0.4-1.6 Pa$\cdot$s at a fixed adhesion energy per unit of area of 0.76  $\mu$J/m$^2$. (b) Experimental data (points) and theoretical predictions (lines) for different adhesion energy per unit area  int he range of 0.69-0.79 $\mu$J/m$^2$ at a fixed membrane viscosity of 0.8 Pa$\cdot$s. c) Time to fully wrap the dumbbell-shaped particle in simulations for a range of tensions <10 nN/m.}
\label{fig:wrapping time vs tension}
\end{figure}

The overall time as well as the transition between different stages in the wrapping strongly depended on the membrane tension - both the initial tension as well as the tension at later times which will increase because of the wrapping, see Figure \ref{fig:wrapping time vs tension}.  We experimentally measured the membrane tension from the fluctuation spectrum of the lipid vesicle following ref. \cite{Pecreaux2004} and plot the time taken to complete wrapping as a function of membrane tension in Figure \ref{fig:wrapping time vs tension}a,b.
We observed an increase in overall wrapping time with increasing initial membrane tension in experiments (e.g. Figure \ref{fig:wrapping time vs tension}a,b) and simulations (Figure \ref{fig:wrapping time vs tension}c). 
However, the range of tensions we could replicate in experiments and simulations was quite limited. To be able to fully explore this effect, we extended a previously developed analytical theory describing the time to wrap colloids \cite{Agudo-Canalejo2015,Frey2019}, which was recently experimentally confirmed \cite{Spanke2022}, and adapted it to the shape of a dumbbell (Details of the theory can be found in the SI). In doing so we could explore the effect of tension on time to wrap the dumbbell for a range of theoretical parameters. All the parameters used in the theory were taken directly from the experiment, apart from the binding energy per area (W) and the microviscosity of the membrane ($\eta_{\mathrm{eff}}$) which are both discussed below.

For a given adhesion energy, we find that the time taken to fully wrap the dumbbell increases non-linearly with the tension. With increasing adhesion energy, the wrapping process becomes faster at the same tension, see Figure \ref{fig:wrapping time vs tension}a. 
The adhesion energies in experiments vary due to the distribution of binding sites between dumbbells \cite{VanDerWel2016, VanDerWel2017a} which is also reflected in that the experimental data points fall within a range of adhesion energies identified by the theory. We note that only a small percentage of the NeutrAvidin sites that have been added during synthesis contribute to the effective adhesion energy, as was found previously in ref. \cite{VanDerWel2016}. Although fixed in the experiments, varying membrane microviscosity in the theory also changes the time taken to wrap. Membrane microviscosity is a measure of how easily the lipids slide past each other during rearrangement, and a higher microviscosity is linked to a higher frictional force during colloid-membrane wrapping. The comparison between the theoretical and experimental results allows us to estimate the membrane microviscosity, which is experimentally inaccessible. We find that our experimental measurements best fit the theoretical curves for a membrane microviscosity of $\eta_{\mathrm{eff}} \approx 0.8 $ Pa$\cdot$s, Figure \ref{fig:wrapping time vs tension}b, about 10 times larger than the lower bound estimated in \cite{Spanke2022}. However, the theory in ref \cite{Spanke2022} consistently over-estimated the wrapping speed as compared with  experiments on spheres, so it could be that the experiment microviscosity was larger than their theoretically predicted value.

Here we have developed the first model system to quantitatively study ligand-receptor mediated endocytosis of an anisotropic object by making use of GUVs and colloidal dumbbell particles. We followed and quantified their orientation $\theta$ and distance $d$ with respect to the membrane during wrapping using experiments and molecular dynamics simulations. We found that there are two final states: 1) only one lobe or 2) both lobes of the dumbbell are fully wrapped by the membrane. The two states can be reached from any initial position except when both lobes were attached initially which necessarily leads to full wrapping of both lobes. However, the initial position influenced the pathway towards the final state. We identified a number of key intermediate states during the wrapping that determine the final state. Wrapping of one lobe was only found for high membrane tensions and if the other lobe did not touch the membrane during engulfment. Using molecular dynamics simulations we quantified the time required between key intermediate steps, with the slowest step being the crossing of the highly curved neck region of the dumbbell. With simulations we confirmed the experimentally-observed trend of time to wrap increasing for increasing tension, and using analytical theory we estimated the membrane microviscosity.

Our results contribute to a better understanding of how shape affects endocytosis, nutrition uptake, and bacterial evasion. 
Our choice of a simple anisotropic object, a dumbbell, enabled a key insight: highly negatively curved regions may dominate the wrapping and possibly even prevent full engulfment unless active processes are present. This suggests that objects, such as certain viruses such as pox virus\cite{schmidt2012poxvirus}that rely on endocytosis, may profit from having a convex shape. 
Incorporation of active processes, such as those driven by actin or ESCRT-III polymers, could provide further insights into how the competition between the passive and active processes affects wrapping.

\subsection*{Supporting Information}
\begin{itemize}
  \item "Supporting Information: Details of the experiments and simulations; experimental materials used; methods employed for membrane preparation, particle functionalization, experimental imaging, and quantification of ligands on dumbbells; details of quantification of dumbbell wrapping and filters applied for the analysis; detail of theory used for time taken to wrap a dumbbell"
  \item Movie S1: An example of an experimental video of the wrapping process of a dumbbell colloid attached to a GUV, recorded with a confocal microscope at 59 frames per second and 3$\times$ accelerated. 

\end{itemize}

Movie S1: An example of an experimental video of the wrapping process of a dumbbell colloid attached to a GUV, recorded with a confocal microscope at 59 frames per second and 3$\times$ accelerated. 

\subsection*{Acknowledgments}
We sincerely thank Casper van der Wel for providing open-source packages for tracking, as well as Yogesh Shelke for his assistance with PAA coverslip preparation and Rachel Doherty for her assistance with particle functionalization. We are grateful to Felix Frey for useful discussions on the theory of membrane wrapping. B.M. and A.Š. acknowledge funding by the European Union’s Horizon 2020 research and innovation programme (ERC Starting Grant No.~802960).

\bibliography{references}

\providecommand{\latin}[1]{#1}
\makeatletter
\providecommand{\doi}
  {\begingroup\let\do\@makeother\dospecials
  \catcode`\{=1 \catcode`\}=2 \doi@aux}
\providecommand{\doi@aux}[1]{\endgroup\texttt{#1}}
\makeatother
\providecommand*\mcitethebibliography{\thebibliography}
\csname @ifundefined\endcsname{endmcitethebibliography}
  {\let\endmcitethebibliography\endthebibliography}{}
\begin{mcitethebibliography}{41}
\providecommand*\natexlab[1]{#1}
\providecommand*\mciteSetBstSublistMode[1]{}
\providecommand*\mciteSetBstMaxWidthForm[2]{}
\providecommand*\mciteBstWouldAddEndPuncttrue
  {\def\EndOfBibitem{\unskip.}}
\providecommand*\mciteBstWouldAddEndPunctfalse
  {\let\EndOfBibitem\relax}
\providecommand*\mciteSetBstMidEndSepPunct[3]{}
\providecommand*\mciteSetBstSublistLabelBeginEnd[3]{}
\providecommand*\EndOfBibitem{}
\mciteSetBstSublistMode{f}
\mciteSetBstMaxWidthForm{subitem}{(\alph{mcitesubitemcount})}
\mciteSetBstSublistLabelBeginEnd
  {\mcitemaxwidthsubitemform\space}
  {\relax}
  {\relax}

\bibitem[Douglas and Young(2006)Douglas, and Young]{Douglas2006}
Douglas,~T.; Young,~M. {Viruses: Making friends with old foes}. \emph{Science}
  \textbf{2006}, \emph{312}, 873--875\relax
\mciteBstWouldAddEndPuncttrue
\mciteSetBstMidEndSepPunct{\mcitedefaultmidpunct}
{\mcitedefaultendpunct}{\mcitedefaultseppunct}\relax
\EndOfBibitem
\bibitem[Manchester and Singh(2006)Manchester, and Singh]{Manchester2006}
Manchester,~M.; Singh,~P. {Virus-based nanoparticles (VNPs): Platform
  technologies for diagnostic imaging}. \emph{Advanced Drug Delivery Reviews}
  \textbf{2006}, \emph{58}, 1505--1522\relax
\mciteBstWouldAddEndPuncttrue
\mciteSetBstMidEndSepPunct{\mcitedefaultmidpunct}
{\mcitedefaultendpunct}{\mcitedefaultseppunct}\relax
\EndOfBibitem
\bibitem[Lewis \latin{et~al.}(2006)Lewis, Destito, Zijlstra, Gonzalez, Quigley,
  Manchester, and Stuhlmann]{Lewis2006}
Lewis,~J.~D.; Destito,~G.; Zijlstra,~A.; Gonzalez,~M.~J.; Quigley,~J.~P.;
  Manchester,~M.; Stuhlmann,~H. {Viral nanoparticles as tools for intravital
  vascular imaging}. \emph{Nature Medicine} \textbf{2006}, \emph{12},
  354--360\relax
\mciteBstWouldAddEndPuncttrue
\mciteSetBstMidEndSepPunct{\mcitedefaultmidpunct}
{\mcitedefaultendpunct}{\mcitedefaultseppunct}\relax
\EndOfBibitem
\bibitem[Schmidt \latin{et~al.}(2012)Schmidt, Bleck, and
  Mercer]{schmidt2012poxvirus}
Schmidt,~F.~I.; Bleck,~C. K.~E.; Mercer,~J. Poxvirus host cell entry.
  \emph{Current opinion in virology} \textbf{2012}, \emph{2}, 20--27\relax
\mciteBstWouldAddEndPuncttrue
\mciteSetBstMidEndSepPunct{\mcitedefaultmidpunct}
{\mcitedefaultendpunct}{\mcitedefaultseppunct}\relax
\EndOfBibitem
\bibitem[Mitchell \latin{et~al.}(2021)Mitchell, Billingsley, Haley, Wechsler,
  Peppas, and Langer]{mitchell_engineering_2021}
Mitchell,~M.~J.; Billingsley,~M.~M.; Haley,~R.~M.; Wechsler,~M.~E.;
  Peppas,~N.~A.; Langer,~R. Engineering precision nanoparticles for drug
  delivery. \emph{Nature Reviews Drug Discovery} \textbf{2021}, \emph{20},
  101--124, Number: 2 Publisher: Nature Publishing Group\relax
\mciteBstWouldAddEndPuncttrue
\mciteSetBstMidEndSepPunct{\mcitedefaultmidpunct}
{\mcitedefaultendpunct}{\mcitedefaultseppunct}\relax
\EndOfBibitem
\bibitem[Richards and Endres(2014)Richards, and Endres]{Richards2014}
Richards,~D.~M.; Endres,~R.~G. {The mechanism of phagocytosis: Two stages of
  engulfment}. \emph{Biophysical Journal} \textbf{2014}, \emph{107},
  1542--1553\relax
\mciteBstWouldAddEndPuncttrue
\mciteSetBstMidEndSepPunct{\mcitedefaultmidpunct}
{\mcitedefaultendpunct}{\mcitedefaultseppunct}\relax
\EndOfBibitem
\bibitem[Spanke \latin{et~al.}(2020)Spanke, Style, Fran{\c{c}}ois-Martin,
  Feofilova, Eisentraut, Kress, Agudo-Canalejo, and Dufresne]{Spanke2020}
Spanke,~H.~T.; Style,~R.~W.; Fran{\c{c}}ois-Martin,~C.; Feofilova,~M.;
  Eisentraut,~M.; Kress,~H.; Agudo-Canalejo,~J.; Dufresne,~E.~R. {Wrapping of
  Microparticles by Floppy Lipid Vesicles}. \emph{Physical Review Letters}
  \textbf{2020}, \emph{125}, 1--9\relax
\mciteBstWouldAddEndPuncttrue
\mciteSetBstMidEndSepPunct{\mcitedefaultmidpunct}
{\mcitedefaultendpunct}{\mcitedefaultseppunct}\relax
\EndOfBibitem
\bibitem[Spanke \latin{et~al.}(2022)Spanke, Agudo-Canalejo, Tran, Style, and
  Dufresne]{Spanke2022}
Spanke,~H.~T.; Agudo-Canalejo,~J.; Tran,~D.; Style,~R.~W.; Dufresne,~E.~R.
  {Dynamics of spontaneous wrapping of microparticles by floppy lipid
  membranes}. \emph{Physical Review Research} \textbf{2022}, \emph{4},
  23080\relax
\mciteBstWouldAddEndPuncttrue
\mciteSetBstMidEndSepPunct{\mcitedefaultmidpunct}
{\mcitedefaultendpunct}{\mcitedefaultseppunct}\relax
\EndOfBibitem
\bibitem[Young(2006)]{young_selective_2006}
Young,~K.~D. The selective value of bacterial shape. \emph{Microbiology and
  molecular biology reviews} \textbf{2006}, \emph{70}, 660--703\relax
\mciteBstWouldAddEndPuncttrue
\mciteSetBstMidEndSepPunct{\mcitedefaultmidpunct}
{\mcitedefaultendpunct}{\mcitedefaultseppunct}\relax
\EndOfBibitem
\bibitem[Levy \latin{et~al.}(1994)Levy, Fraenkel-Conrat, and
  Owens]{levy_virology_1994}
Levy,~J.~A.; Fraenkel-Conrat,~H.; Owens,~O.~S. \emph{Virology}, 3rd ed.;
  Benjamin Cummings: Englewood Cliffs, N.J, 1994\relax
\mciteBstWouldAddEndPuncttrue
\mciteSetBstMidEndSepPunct{\mcitedefaultmidpunct}
{\mcitedefaultendpunct}{\mcitedefaultseppunct}\relax
\EndOfBibitem
\bibitem[Cho \latin{et~al.}(2008)Cho, Wang, Nie, Chen, and
  Shin]{cho_therapeutic_2008}
Cho,~K.; Wang,~X.; Nie,~S.; Chen,~Z.~G.; Shin,~D.~M. Therapeutic
  {Nanoparticles} for {Drug} {Delivery} in {Cancer}. \emph{Clinical Cancer
  Research} \textbf{2008}, \emph{14}, 1310--1316\relax
\mciteBstWouldAddEndPuncttrue
\mciteSetBstMidEndSepPunct{\mcitedefaultmidpunct}
{\mcitedefaultendpunct}{\mcitedefaultseppunct}\relax
\EndOfBibitem
\bibitem[Aoyama \latin{et~al.}(2003)Aoyama, Kanamori, Nakai, Sasaki, Horiuchi,
  Sando, and Niidome]{Aoyama2003}
Aoyama,~Y.; Kanamori,~T.; Nakai,~T.; Sasaki,~T.; Horiuchi,~S.; Sando,~S.;
  Niidome,~T. {Artificial viruses and their application to gene delivery.
  Size-controlled gene coating with glycocluster nanoparticles}. \emph{Journal
  of the American Chemical Society} \textbf{2003}, \emph{125}, 3455--3457\relax
\mciteBstWouldAddEndPuncttrue
\mciteSetBstMidEndSepPunct{\mcitedefaultmidpunct}
{\mcitedefaultendpunct}{\mcitedefaultseppunct}\relax
\EndOfBibitem
\bibitem[Richards and Endres(2016)Richards, and Endres]{Richards2016}
Richards,~D.~M.; Endres,~R.~G. {Target shape dependence in a simple model of
  receptor-mediated endocytosis and phagocytosis}. \emph{Proceedings of the
  National Academy of Sciences of the United States of America} \textbf{2016},
  \emph{113}, 6113--6118\relax
\mciteBstWouldAddEndPuncttrue
\mciteSetBstMidEndSepPunct{\mcitedefaultmidpunct}
{\mcitedefaultendpunct}{\mcitedefaultseppunct}\relax
\EndOfBibitem
\bibitem[Chithrani \latin{et~al.}(2006)Chithrani, Ghazani, and
  Chan]{Chithrani2006}
Chithrani,~B.~D.; Ghazani,~A.~A.; Chan,~W.~C. {Determining the size and shape
  dependence of gold nanoparticle uptake into mammalian cells}. \emph{Nano
  Letters} \textbf{2006}, \emph{6}, 662--668\relax
\mciteBstWouldAddEndPuncttrue
\mciteSetBstMidEndSepPunct{\mcitedefaultmidpunct}
{\mcitedefaultendpunct}{\mcitedefaultseppunct}\relax
\EndOfBibitem
\bibitem[Chithrani and Chan(2007)Chithrani, and Chan]{Chithrani2007}
Chithrani,~B.~D.; Chan,~W.~C. {Elucidating the mechanism of cellular uptake and
  removal of protein-coated gold nanoparticles of different sizes and shapes}.
  \emph{Nano Letters} \textbf{2007}, \emph{7}, 1542--1550\relax
\mciteBstWouldAddEndPuncttrue
\mciteSetBstMidEndSepPunct{\mcitedefaultmidpunct}
{\mcitedefaultendpunct}{\mcitedefaultseppunct}\relax
\EndOfBibitem
\bibitem[Li(2012)]{Li2013}
Li,~X. {Size and shape effects on receptor-mediated endocytosis of
  nanoparticles}. \emph{Journal of Applied Physics} \textbf{2012}, \emph{111},
  2010--2014\relax
\mciteBstWouldAddEndPuncttrue
\mciteSetBstMidEndSepPunct{\mcitedefaultmidpunct}
{\mcitedefaultendpunct}{\mcitedefaultseppunct}\relax
\EndOfBibitem
\bibitem[Dasgupta and Dimova(2014)Dasgupta, and Dimova]{Dasgupta2014}
Dasgupta,~R.; Dimova,~R. {Inward and outward membrane tubes pulled from giant
  vesicles}. \emph{Journal of Physics D: Applied Physics} \textbf{2014},
  \emph{47}, 282001\relax
\mciteBstWouldAddEndPuncttrue
\mciteSetBstMidEndSepPunct{\mcitedefaultmidpunct}
{\mcitedefaultendpunct}{\mcitedefaultseppunct}\relax
\EndOfBibitem
\bibitem[Van Der~Wel \latin{et~al.}(2016)Van Der~Wel, Vahid, {\v{S}}ari{\'{C}},
  Idema, Heinrich, and Kraft]{VanDerWel2016}
Van Der~Wel,~C.; Vahid,~A.; {\v{S}}ari{\'{C}},~A.; Idema,~T.; Heinrich,~D.;
  Kraft,~D.~J. {Lipid membrane-mediated attraction between curvature inducing
  objects}. \emph{Scientific Reports} \textbf{2016}, \emph{6}, 1--10\relax
\mciteBstWouldAddEndPuncttrue
\mciteSetBstMidEndSepPunct{\mcitedefaultmidpunct}
{\mcitedefaultendpunct}{\mcitedefaultseppunct}\relax
\EndOfBibitem
\bibitem[Decuzzi and Ferrari(2008)Decuzzi, and
  Ferrari]{decuzzi_receptor-mediated_2008}
Decuzzi,~P.; Ferrari,~M. The {Receptor}-{Mediated} {Endocytosis} of
  {Nonspherical} {Particles}. \emph{Biophysical Journal} \textbf{2008},
  \emph{94}, 3790--3797\relax
\mciteBstWouldAddEndPuncttrue
\mciteSetBstMidEndSepPunct{\mcitedefaultmidpunct}
{\mcitedefaultendpunct}{\mcitedefaultseppunct}\relax
\EndOfBibitem
\bibitem[V{\'{a}}cha \latin{et~al.}(2011)V{\'{a}}cha, Martinez-Veracoechea, and
  Frenkel]{Vacha2011}
V{\'{a}}cha,~R.; Martinez-Veracoechea,~F.~J.; Frenkel,~D. {Receptor-mediated
  endocytosis of nanoparticles of various shapes}. \emph{Nano Letters}
  \textbf{2011}, \emph{11}, 5391--5395\relax
\mciteBstWouldAddEndPuncttrue
\mciteSetBstMidEndSepPunct{\mcitedefaultmidpunct}
{\mcitedefaultendpunct}{\mcitedefaultseppunct}\relax
\EndOfBibitem
\bibitem[Bahrami(2013)]{Bahrami2013OrientationalMembranes}
Bahrami,~A.~H. {Orientational changes and impaired internalization of
  ellipsoidal nanoparticles by vesicle membranes}. \emph{Soft Matter}
  \textbf{2013}, \emph{9}, 8642--8646\relax
\mciteBstWouldAddEndPuncttrue
\mciteSetBstMidEndSepPunct{\mcitedefaultmidpunct}
{\mcitedefaultendpunct}{\mcitedefaultseppunct}\relax
\EndOfBibitem
\bibitem[Yang \latin{et~al.}(2013)Yang, Yuan, and Ma]{yang_influence_2013}
Yang,~K.; Yuan,~B.; Ma,~Y.-q. Influence of geometric nanoparticle rotation on
  cellular internalization process. \emph{Nanoscale} \textbf{2013}, \emph{5},
  7998--8006, Publisher: The Royal Society of Chemistry\relax
\mciteBstWouldAddEndPuncttrue
\mciteSetBstMidEndSepPunct{\mcitedefaultmidpunct}
{\mcitedefaultendpunct}{\mcitedefaultseppunct}\relax
\EndOfBibitem
\bibitem[Huang \latin{et~al.}(2013)Huang, Zhang, Yuan, Gao, and
  Zhang]{Huang2013}
Huang,~C.; Zhang,~Y.; Yuan,~H.; Gao,~H.; Zhang,~S. {Role of nanoparticle
  geometry in endocytosis: Laying down to stand up}. \emph{Nano Letters}
  \textbf{2013}, \emph{13}, 4546--4550\relax
\mciteBstWouldAddEndPuncttrue
\mciteSetBstMidEndSepPunct{\mcitedefaultmidpunct}
{\mcitedefaultendpunct}{\mcitedefaultseppunct}\relax
\EndOfBibitem
\bibitem[Dasgupta \latin{et~al.}(2014)Dasgupta, Auth, and
  Gompper]{dasgupta_shape_2014}
Dasgupta,~S.; Auth,~T.; Gompper,~G. Shape and {Orientation} {Matter} for the
  {Cellular} {Uptake} of {Nonspherical} {Particles}. \emph{Nano Letters}
  \textbf{2014}, \emph{14}, 687--693, Publisher: American Chemical
  Society\relax
\mciteBstWouldAddEndPuncttrue
\mciteSetBstMidEndSepPunct{\mcitedefaultmidpunct}
{\mcitedefaultendpunct}{\mcitedefaultseppunct}\relax
\EndOfBibitem
\bibitem[Bahrami \latin{et~al.}(2014)Bahrami, Raatz, Agudo-Canalejo, Michel,
  Curtis, Hall, Gradzielski, Lipowsky, and Weikl]{Bahrami2014}
Bahrami,~A.~H.; Raatz,~M.; Agudo-Canalejo,~J.; Michel,~R.; Curtis,~E.~M.;
  Hall,~C.~K.; Gradzielski,~M.; Lipowsky,~R.; Weikl,~T.~R. {Wrapping of
  nanoparticles by membranes}. \emph{Advances in Colloid and Interface Science}
  \textbf{2014}, \emph{208}, 214--224\relax
\mciteBstWouldAddEndPuncttrue
\mciteSetBstMidEndSepPunct{\mcitedefaultmidpunct}
{\mcitedefaultendpunct}{\mcitedefaultseppunct}\relax
\EndOfBibitem
\bibitem[Tang \latin{et~al.}(2018)Tang, Zhang, Ye, and Zheng]{Tang2018}
Tang,~H.; Zhang,~H.; Ye,~H.; Zheng,~Y. {Receptor-Mediated Endocytosis of
  Nanoparticles: Roles of Shapes, Orientations, and Rotations of
  Nanoparticles}. \emph{Journal of Physical Chemistry B} \textbf{2018},
  \emph{122}, 171--180\relax
\mciteBstWouldAddEndPuncttrue
\mciteSetBstMidEndSepPunct{\mcitedefaultmidpunct}
{\mcitedefaultendpunct}{\mcitedefaultseppunct}\relax
\EndOfBibitem
\bibitem[Agudo-Canalejo(2020)]{Agudo-Canalejo2020}
Agudo-Canalejo,~J. {Engulfment of ellipsoidal nanoparticles by membranes: full
  description of orientational changes}. \emph{Journal of Physics: Condensed
  Matter} \textbf{2020}, \emph{32}, 294001\relax
\mciteBstWouldAddEndPuncttrue
\mciteSetBstMidEndSepPunct{\mcitedefaultmidpunct}
{\mcitedefaultendpunct}{\mcitedefaultseppunct}\relax
\EndOfBibitem
\bibitem[Champion and Mitragotri(2006)Champion, and Mitragotri]{Champion2006}
Champion,~J.~A.; Mitragotri,~S. {Role of target geometry in phagocytosis}.
  \emph{Proceedings of the National Academy of Sciences of the United States of
  America} \textbf{2006}, \emph{103}, 4930--4934\relax
\mciteBstWouldAddEndPuncttrue
\mciteSetBstMidEndSepPunct{\mcitedefaultmidpunct}
{\mcitedefaultendpunct}{\mcitedefaultseppunct}\relax
\EndOfBibitem
\bibitem[van~der Wel \latin{et~al.}(2017)van~der Wel, Heinrich, and
  Kraft]{VanDerWel2017}
van~der Wel,~C.; Heinrich,~D.; Kraft,~D.~J. {Microparticle Assembly Pathways on
  Lipid Membranes}. \emph{Biophysical Journal} \textbf{2017}, \emph{113},
  1037--1046\relax
\mciteBstWouldAddEndPuncttrue
\mciteSetBstMidEndSepPunct{\mcitedefaultmidpunct}
{\mcitedefaultendpunct}{\mcitedefaultseppunct}\relax
\EndOfBibitem
\bibitem[Sarfati and Dufresne(2016)Sarfati, and Dufresne]{Sarfati2016}
Sarfati,~R.; Dufresne,~E.~R. {Long-range attraction of particles adhered to
  lipid vesicles}. \emph{Physical Review E} \textbf{2016}, \emph{94},
  2--7\relax
\mciteBstWouldAddEndPuncttrue
\mciteSetBstMidEndSepPunct{\mcitedefaultmidpunct}
{\mcitedefaultendpunct}{\mcitedefaultseppunct}\relax
\EndOfBibitem
\bibitem[Van Der~Wel \latin{et~al.}(2017)Van Der~Wel, Bossert, Mank, Winter,
  Heinrich, and Kraft]{VanDerWel2017a}
Van Der~Wel,~C.; Bossert,~N.; Mank,~Q.~J.; Winter,~M.~G.; Heinrich,~D.;
  Kraft,~D.~J. {Surfactant-free Colloidal Particles with Specific Binding
  Affinity}. \emph{Langmuir} \textbf{2017}, \emph{33}, 9803--9810\relax
\mciteBstWouldAddEndPuncttrue
\mciteSetBstMidEndSepPunct{\mcitedefaultmidpunct}
{\mcitedefaultendpunct}{\mcitedefaultseppunct}\relax
\EndOfBibitem
\bibitem[Meester \latin{et~al.}(2016)Meester, Verweij, Van Der~Wel, and
  Kraft]{Meester2016ColloidalParticles}
Meester,~V.; Verweij,~R.~W.; Van Der~Wel,~C.; Kraft,~D.~J. {Colloidal
  Recycling: Reconfiguration of Random Aggregates into Patchy Particles}.
  \emph{ACS Nano} \textbf{2016}, \emph{10}, 4322--4329\relax
\mciteBstWouldAddEndPuncttrue
\mciteSetBstMidEndSepPunct{\mcitedefaultmidpunct}
{\mcitedefaultendpunct}{\mcitedefaultseppunct}\relax
\EndOfBibitem
\bibitem[Dietrich \latin{et~al.}(1997)Dietrich, Angelova, and
  Pouligny]{Dietrich1997}
Dietrich,~C.; Angelova,~M.; Pouligny,~B. {Adhesion of Latex spheres to giant
  phospholipid vesicles: Statics and dynamics}. \emph{Journal de physique. II}
  \textbf{1997}, \emph{7}, 1651--1682\relax
\mciteBstWouldAddEndPuncttrue
\mciteSetBstMidEndSepPunct{\mcitedefaultmidpunct}
{\mcitedefaultendpunct}{\mcitedefaultseppunct}\relax
\EndOfBibitem
\bibitem[Yuan \latin{et~al.}(2010)Yuan, Huang, Li, Lykotrafitis, and
  Zhang]{Yuan2010}
Yuan,~H.; Huang,~C.; Li,~J.; Lykotrafitis,~G.; Zhang,~S. One-particle-thick,
  solvent-free, coarse-grained model for biological and biomimetic fluid
  membranes. \emph{Phys. Rev. E} \textbf{2010}, \emph{82}, 011905\relax
\mciteBstWouldAddEndPuncttrue
\mciteSetBstMidEndSepPunct{\mcitedefaultmidpunct}
{\mcitedefaultendpunct}{\mcitedefaultseppunct}\relax
\EndOfBibitem
\bibitem[Curk \latin{et~al.}(2018)Curk, Wirnsberger, Dobnikar, Frenkel, and
  Šarić]{curk2018controlling}
Curk,~T.; Wirnsberger,~P.; Dobnikar,~J.; Frenkel,~D.; Šarić,~A. Controlling
  cargo trafficking in multicomponent membranes. \emph{Nano letters}
  \textbf{2018}, \emph{18}, 5350--5356\relax
\mciteBstWouldAddEndPuncttrue
\mciteSetBstMidEndSepPunct{\mcitedefaultmidpunct}
{\mcitedefaultendpunct}{\mcitedefaultseppunct}\relax
\EndOfBibitem
\bibitem[Vanhille-Campos and {\v{S}}ari{\'c}(2021)Vanhille-Campos, and
  {\v{S}}ari{\'c}]{vanhille2021modelling}
Vanhille-Campos,~C.; {\v{S}}ari{\'c},~A. Modelling the dynamics of vesicle
  reshaping and scission under osmotic shocks. \emph{Soft Matter}
  \textbf{2021}, \emph{17}, 3798--3806\relax
\mciteBstWouldAddEndPuncttrue
\mciteSetBstMidEndSepPunct{\mcitedefaultmidpunct}
{\mcitedefaultendpunct}{\mcitedefaultseppunct}\relax
\EndOfBibitem
\bibitem[{\v{S}}ari{\'c} and Cacciuto(2012){\v{S}}ari{\'c}, and
  Cacciuto]{vsaric2012mechanism}
{\v{S}}ari{\'c},~A.; Cacciuto,~A. Mechanism of membrane tube formation induced
  by adhesive nanocomponents. \emph{Physical review letters} \textbf{2012},
  \emph{109}, 188101\relax
\mciteBstWouldAddEndPuncttrue
\mciteSetBstMidEndSepPunct{\mcitedefaultmidpunct}
{\mcitedefaultendpunct}{\mcitedefaultseppunct}\relax
\EndOfBibitem
\bibitem[P{\'{e}}cr{\'{e}}aux \latin{et~al.}(2004)P{\'{e}}cr{\'{e}}aux,
  D{\"{o}}bereiner, Prost, Joanny, and Bassereau]{Pecreaux2004}
P{\'{e}}cr{\'{e}}aux,~J.; D{\"{o}}bereiner,~H.~G.; Prost,~J.; Joanny,~J.~F.;
  Bassereau,~P. {Refined contour analysis of giant unilamellar vesicles}.
  \emph{European Physical Journal E} \textbf{2004}, \emph{13}, 277--290\relax
\mciteBstWouldAddEndPuncttrue
\mciteSetBstMidEndSepPunct{\mcitedefaultmidpunct}
{\mcitedefaultendpunct}{\mcitedefaultseppunct}\relax
\EndOfBibitem
\bibitem[Agudo-Canalejo and Lipowsky(2015)Agudo-Canalejo, and
  Lipowsky]{Agudo-Canalejo2015}
Agudo-Canalejo,~J.; Lipowsky,~R. {Critical particle sizes for the engulfment of
  nanoparticles by membranes and vesicles with bilayer asymmetry}. \emph{ACS
  Nano} \textbf{2015}, \emph{9}, 3704--3720\relax
\mciteBstWouldAddEndPuncttrue
\mciteSetBstMidEndSepPunct{\mcitedefaultmidpunct}
{\mcitedefaultendpunct}{\mcitedefaultseppunct}\relax
\EndOfBibitem
\bibitem[Frey \latin{et~al.}(2019)Frey, Ziebert, and Schwarz]{Frey2019}
Frey,~F.; Ziebert,~F.; Schwarz,~U.~S. {Stochastic dynamics of nanoparticle and
  virus uptake}. \emph{Physical Review Letters} \textbf{2019}, \emph{122},
  88102\relax
\mciteBstWouldAddEndPuncttrue
\mciteSetBstMidEndSepPunct{\mcitedefaultmidpunct}
{\mcitedefaultendpunct}{\mcitedefaultseppunct}\relax
\EndOfBibitem
\end{mcitethebibliography}

\end{document}